\documentclass[9pt]{article}
\usepackage{amsfonts}
\usepackage{spconf,amsmath,graphicx}
\usepackage{cite}
\graphicspath{{Figs/}}
\usepackage{float}
\usepackage{graphicx}
\usepackage{epsf}
\usepackage{array}
\usepackage{makecell}
\usepackage{amssymb}
\usepackage{lineno}
\usepackage{fixmath}
\usepackage{hyperref}
\usepackage{float}

\title{Deep Residual Echo Suppression with A Tunable Tradeoff Between Signal Distortion and Echo Suppression}
\name{Amir Ivry \qquad Israel Cohen \qquad Baruch Berdugo \thanks{This work was supported by the Israel Science Foundation (grant no. 576/16) and the ISF-NSFC joint research program (grant No. 2514/17).}}
\address{Andrew and Erna Viterbi Faculty of Electrical Engineering \\
	Technion -- Israel Institute of Technology, Technion City, Haifa 3200003, Israel}
\begin{document}
%
\maketitle
\begin{abstract}
In this paper, we propose a residual echo suppression method using a UNet neural network that directly maps the outputs of a linear acoustic echo canceler to the desired signal in the spectral domain. This system embeds a design parameter that allows a tunable tradeoff between the desired-signal distortion and residual echo suppression in double-talk scenarios. The system employs 136 thousand parameters, and requires 1.6 Giga floating-point operations per second and 10 Mega-bytes of memory. The implementation satisfies both the timing requirements of the AEC challenge and the computational and memory limitations of on-device applications. Experiments are conducted with 161~h of data from the AEC challenge database and from real independent recordings. We demonstrate the performance of the proposed system in real-life conditions and compare it with two competing methods regarding echo suppression and desired-signal distortion, generalization to various environments, and robustness to high echo levels.
\end{abstract}
\begin{keywords}
Residual echo suppression, on-device implementation, acoustic echo cancellation, UNet.
\end{keywords}
\section{Introduction}
\label{sec:intro}
Real-life telecommunication scenarios involve a conversation between two speakers that are located at near-end and far-end points. The near-end includes a microphone that captures the near-end signal, echo produced by a loudspeaker playing the far-end signal, and background noises \cite{sondhi1995stereophonic}. The presence of acoustic echo can lead to degradation in intelligibility and quality of conversation, since the far-end speaker can hear their own voice while speaking, and near-end speech can be screened. Conventional acoustic echo cancelers (AECs) do not model non-linearities in the echo path, and generally introduce a mismatch between true and estimated echo paths during convergence and re-convergence \cite{benesty1998better}. This results in residual echo that must be suppressed by a dedicated system.

Deep learning has occupied a major role in AEC studies and showed enhanced performance compared to traditional methods \cite{halimeh2020efficient}, \cite{fazel2020cad}. A recent study exploited long short-term memory (LSTM) networks to jointly obtain echo cancellation and to suppress noises and reverberations \cite{carbajal2019joint}. Lee et al. \cite{lee2015dnn} cascaded a fully-connected neural network (FCNN) after a linear acoustic echo suppressor (AES) and evaluated the objective gain between the spectra amplitudes of the near-end and AES output signals. Lei et al. \cite{lei2019deep} exploited past and future temporal context to map the microphone and reference far-end signals to the desired speaker via an FCNN. Lately, deep learning and classic methods were jointly utilized in \cite{ma2020acoustic} and \cite{zhang2019deep}, where the latter activated convolutional recurrent networks to evaluate the real and imaginary parts of the near-end signal spectrogram.

In this study, we introduce a residual echo suppression (RES) method with a dual-channel input and single-channel output UNet neural network that directly maps the outputs of a linear AEC to the desired near-end signal in the short-time Fourier transform (STFT) domain. By utilizing the depth-wise separable convolution in every convolution layer of the UNet \cite{gadosey2020sd}, the system comprises 136 thousand parameters that consume 1.6 Giga floating point operations per second (flops) and 10 Mega-bytes (MB) of memory, which makes it suitable for on-device integration. Also, the system meets the timing standards of the AEC challenge \cite{sridhar2020icassp}, and more generally the constraints of hands-free communication systems \cite{handsfree}.

Even though competing models \cite{halimeh2020efficient}--\cite{zhang2019deep}, \cite{zhang2018deep}, \cite{carbajal2018multiple} have shown promising results, the performance in real acoustic environments is still challenging. Furthermore, a tunable tradeoff between the level of RES and desired-signal distortion may benefit applications that vary in their specific tradeoff requirements. However, this feature is not enabled by design in existing approaches. We bridge these gaps as follows. First, we conduct experiments with over 160~h of data that was acquired from the AEC challenge database \cite{sridhar2020icassp} and from independent recordings in real conditions. Second, a design parameter that allows dynamic balance between echo reduction and signal distortion is embedded in the UNet objective function that is minimized during the training process.

The performance of the proposed system is compared to two existing deep learning-based methods: Zhang and Wang \cite{zhang2018deep}, where a bi-LSTM structure was utilized to model an ideal ratio mask for AEC and then for RES, and Carbajal et al. \cite{carbajal2018multiple}, who introduced a multiple input FCNN RES system, fed with linear AEC outputs and a reference far-end signal to estimate a phase-sensitive mask.
Experimental results show state-of-the-art performance of the proposed method in various real-life acoustic setups. Particularly, high generalization is demonstrated in a variety of environments, devices, speakers, and moving echo paths. High robustness is also achieved in extreme conditions of very low signal-to-echo-ratios (SERs), and the effect of the tunable design parameter is demonstrated.

The reminder of this paper is organized as follows. Section \ref{ProblemFormulation} formulates the problem. Section \ref{algorithm} introduces the proposed system. Section \ref{Setup} details the experimental setup. Section \ref{results} reports obtained performance. Section \ref{conclusion} concludes.

\begin{figure}[t]
	\centering\includegraphics[width=3.3in]{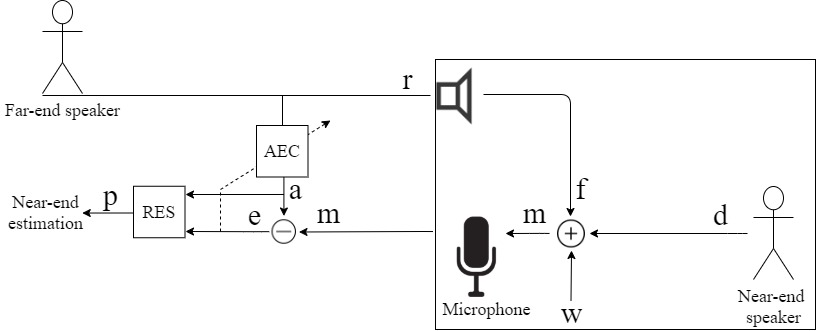}
	\caption{Echo cancellation system. Time indices are neglected.}
	\label{fig: setup}
\end{figure}

\section{Problem Formulation}\label{ProblemFormulation}
Let $r\left[n\right]$ denote the reference far-end signal and let $d\left[n\right]$ denote the desired near-end signal in the discrete time domain $\forall n\in\mathbb{Z^{+}}$. The microphone signal, $m\left[n\right]$, is given by
\begin{align}
m\left[n\right] = f\left[n\right] + d\left[n\right] + w\left[n\right],
\label{eq:micModel}
\end{align}
\noindent where $f\left[n\right]$ is a reverberant non-linear modification of $r\left[n\right]$ and $w\left[n\right]$ denotes environmental and inherent system noises.

Before applying RES, a linear AEC is applied to reduce the linear echo. The AEC receives $m\left[n\right]$ as input and $r\left[n\right]$ as reference, and generates two output signals: $a\left[n\right]$, the outcome of an adaptive filtering process that attempts to model $f\left[n\right]$, and the error signal $e\left[n\right]$ that is given by
\begin{align}
e\left[n\right] = m\left[n\right] - a\left[n\right].
\label{eq:errorModel}
\end{align}
\noindent From (\ref{eq:micModel}) and (\ref{eq:errorModel}) we have
\begin{align}
e\left[n\right] = d\left[n\right] + \left(f\left[n\right]- a\left[n\right]\right) + w\left[n\right].
\label{eq:e}
\end{align}
%
\noindent Namely, $e\left[n\right]$ contains an additive combination of three components: The desired signal $d\left[n\right]$, the noise $w\left[n\right]$, and the residual echo $z\left[n\right]$, given by
\begin{align}
z\left[n\right] = f\left[n\right]- a\left[n\right].
\label{eq:z}
\end{align}

The goal is to suppress the residual echo $z\left[n\right]$ without distorting the desired signal $d\left[n\right]$. Fig. \ref{fig: setup} shows a scheme of the echo cancellation system.

\section{Proposed System}\label{algorithm}
The proposed RES system comprises a UNet neural network with two input channels and one output channel. The network is fed with the STFT amplitude of the linear AEC outputs and aims to recover the STFT amplitude of the desired near-end signal. The contracting and expansive paths of the UNet are each constructed of 5 convolution units. Every unit contains 2 concatenated and identical layers, where every layer consists of 2-D convolution, 2-D batch normalization, and ReLU activation. Here, convolution is implemented in two parts; depth-wise convolution layer with a $3\times3$ kernel and padding of 1, followed by a separable convolution layer, to reduce computational load. During contraction, convolution units are followed by a max pooling layer, and during expansion, convolution units are preceded by an up-sampling layer, both of scaling factor 2. Skip connections are applied between matching pairs of contraction and expansion convolution units.

To exploit the powerful image segmentation abilities of the UNet \cite{gadosey2020sd}, its channels are fed with a long temporal context of 300~ms that generates spectrogram images. During encoding, short filters jointly capture time-frequency local connections and produce numerous features that discriminate residual echo. During decoding, a similar convolution mechanism removes these echo signatures while preserving the desired signal. Long skip connections allow recovery of fine-grained details in the prediction, as features of the same dimension are reemployed from earlier layers, gradient flows directly via skip connections, which enhances optimization, and features are directly passed from encoder to decoder to recover spatial information lost during down-sampling.

A tunable design parameter $\alpha\geq0$ is embedded in a custom loss function $J(\alpha)$ that is minimized during training:
\begin{flalign}
J(\alpha) =
\ell^{2}_{2}\left(P-D\right) + \alpha\,\ell^{2}_{2}\left(P\right)+ 0.1\,\sigma^{2}\left(P\right)\mathbb{I}_{\alpha> 0} \textrm{ ,}
\label{eq: loss}
\end{flalign}
\noindent where $P$ and $D$, respectively, represent the mini-batch predicted and desired spectra amplitudes after normalization, as described in Section \ref{preproc}. $\ell^{2}_{2}$ and $\sigma^{2}$ denote the mean squared $\ell_{2}$-norm and variance operators, and $\mathbb{I}_{\alpha> 0}$ equals 1 when ${\alpha>0}$ and 0 otherwise. During the training stage, $J(\alpha)$ is minimized while $\alpha$ penalizes $\ell^{2}_{2}\left(P\right)$, which allows a dynamic tradeoff between the levels of RES and desired-signal distortion of the system. When $\alpha=0$, the error between the prediction and the near-end signal is minimized. However, when $\alpha>0$, smaller prediction values are generated. This reduces the level of residual echo but compromises the level of desired-signal distortion. $\sigma^{2}\left(P\right)$ mitigates sub-band nullification that may occur when $\alpha\neq0$. A practical usage of $\alpha$ is a tunable user interface parameter for adjusting the performance of the system according to specific user preferences.

The linear AEC system that precedes the UNet was made by Phoenix Audio Technologies and operates based on filter banks. It employs a 150 ms filter length, converges after 1 s, and consumes 200 Kflops. Overall, the joint system is comprised of the AEC and RES contains 136 thousand parameters that consume 1.6 Gflops and memory of 10 MB. This system meets timing constrains of hands-free communication \cite{sridhar2020icassp}, \cite{handsfree} on the standard Intel Core i7-8700K CPU @ 3.7 GHz. Thus, on-device system integration is enabled, e.g., on the AM5749\textsuperscript{TM} processor by Texas Instruments \cite{TI}.

\section{Experimental Setup}\label{Setup}
\subsection{Database Acquisition}\label{database}
The SER and signal-to-noise-ratio (SNR) levels captured by the microphone are calculated by SER~=$10\log_{10}\left[\Vert d \Vert_{2}^{2} / \Vert f \Vert_{2}^{2}\right]$ and SNR~=~$10\log_{10}\left[\Vert d \Vert_{2}^{2} / \Vert w \Vert_{2}^{2}\right]$ in decibels. Both measures are obtained using 50\% overlapping time frames of 20 ms.
Two data corpora were employed in this study; the AEC challenge database \cite{sridhar2020icassp} used for training, and an independently recorded database used for both training and testing.

The AEC challenge database contains two new open sources of synthetic and real recordings. The synthetic data captures $100$ hours of clean and noisy single talk and double talk periods. The real data was derived by a crowd sourcing effort that yielded $50$ hours of audio clips, generated from $2,500$ real acoustic environments, audio devices, and human speaking in single and double talk scenarios that included changed and unchanged echo paths. SER levels were uniformly distributed between -10 and 10 dB and SNR was randomly sampled between 0 and 40 dB.

Also, independent recordings in real-life conditions were conducted to test the generalization of the system to unseen setups and its robustness to low levels of SERs. The near-end signal was generated via a mouth simulator type 4227-A\textsuperscript{TM} of Br\"uel\&Kjaer so its recordings contained inherent and environmental system noises. The microphone and loudspeaker were either enclosed within a distance of 5 cm by speakerphones of type Spider MT503\textsuperscript{TM} or Quattro MT301\textsuperscript{TM}, or the echo was played externally by Logitech type Z120\textsuperscript{TM} loudspeaker. The mouth simulator was placed in three positions located either at 1, 1.5, or 2~m from the microphone, and was shifted only between recordings. Transitions in the echo path were generated by moving the external loudspeaker either 1, 1.5, or 2~m away from the microphone during recordings, producing 3 source-receiver positions. The data used for experiments was equally mixed between 5.5~h from the TIMIT \cite{TIMIT_correct} and 5.5~h from the LibriSpeech \cite{panayotov2015librispeech} corpora. Recordings were performed in 4 different room sizes varied between a $3\times3\times2.5 \textrm{ m}^{3}$ volume to a larger $5\times5\times4 \textrm{ m}^{3}$ volume, and the reverberation time, i.e. $\textrm{RT}_{60}$, varied between 0.3-0.6~s. For double talk utterances, near-end and far-end speakers were chosen randomly, zero-padded to the same length, and added in various SER levels between -10 and -20 dB. The average overlap between near-end and far-end signals was 90\%. The number of far-end single-talk, near-end single-talk, and double-talk utterances was identical. Male and female speakers equally participated, double-talk periods contained two different speakers, the training and test sets did not share the same speakers, and every speaker was both the far-end and near-end speaker. Overall, 11~h of data were generated and equally split between the training and test sets so both contained disjoint and balanced setups in terms of acoustic environments, devices, and speakers. SNR level was $32\pm5$ dB and sample frequency was 16~KHz.

\subsection{Data Processing, Training, and Testing}\label{preproc}
The microphone and reference signals are processed with 50\% overlapping time frames of 20 ms. First, these frames are inserted to the linear AEC. Then, each of the two output frames is represented by 161 frequency bins by taking the amplitude of a 320-point STFT. In training, this spectral data is typically normalized between 0 and 1, i.e., for every frequency bin between 1 and 161, the corresponding vector of frame samples is reduced by its minimum value and divided by its dynamic range. These training statistics are reapplied to the test data. Next, batches of 30 frames without overlap, corresponding to 300 ms, are inserted to both input channels and to the single output channel of the UNet. Training optimization is done by minimizing the loss function in eq. (\ref{eq: loss}) with a learning rate of 0.0005, mini-batch size of 4, and 20 epochs using Adam optimizer \cite{adam_correct}. Training duration was 1.5 hours per 10 hours of training data on an Intel Core i7-8700K CPU @ 3.7 GHz with two GPUs of type Nvidia GeForce RTX 2080 Ti. During testing, normalized batches of 30 frames are inserted to the UNet with a step size of one frame. After the amplitude spectral prediction is generated, every frequency bin undergoes the inverse normalization process described above using the training statistics. This result undergoes an inverse STFT using the error signal phase with the overlap-add method \cite{george1997speech}. An artificial gain may be introduced by the RES and is compensated as shown in \cite{carbajal2018multiple}.
\begin{table}
	\centering
	\caption{Performance Measures for RES.}
	\label{table:metrics}
	\begin{tabular}{|c c|}
		\hline
		Metric & Definition \\ [0.5ex]
		\hline\hline
		ERLE &
		$10\log_{10}\frac{\Vert e \Vert_{2}^2}{\Vert p \Vert_{2}^2} \Bigr|_{\textrm{far-end single talk}}$  \\ [1ex]
		\hline
		SAR &
		$10\log_{10}\frac{\Vert d \Vert_{2}^2}{\Vert p-d \Vert_{2}^2}\Bigr|_{\textrm{near-end single talk}}$ \\
		\hline
		SDR &
		$10\log_{10}\frac{\Vert d \Vert_{2}^2}{\Vert p-d \Vert_{2}^2}\Bigr|_{\textrm{double talk scenario}}$ \\
		\hline
	\end{tabular}
\end{table}
\subsection{Performance Measures}\label{measures}
To evaluate performance we use the echo return loss enhancement (ERLE) \cite{ERLE} that measures the echo reduction between the noisy and enhanced signals when only far-end signal is present, and signal-to-artifacts-ratio (SAR) that measures the distortion for near-end single-talk periods \cite{vincent2006performance}. For double-talk periods, we use the signal-to-distortion-ratio (SDR) \cite{vincent2006performance} that takes echo suppression and speech artifacts into account, and the perceptual evaluation of speech quality (PESQ) \cite{PESQ}. The performance measures are defined in Table~\ref{table:metrics}. Besides the PESQ that is calculated over an entire utterance, these measures are calculated with 50\% overlapping frames of 20~ms.

\section{Experimental Results}\label{results}
We compare the performance of the proposed system with two competing deep learning-based RES methods in \cite{zhang2018deep}, referring to its reported ``AES+BLSTM'' system, and \cite{carbajal2018multiple}. All RES models are fed with the outputs of the same linear AEC discussed in this study. In all experiments, the linear AEC has converged and $\alpha=0$ unless stated otherwise. Every model is trained using both the entire AEC challenge data and independently recorded training data, which accumulates to over 155~h. Performance measures are reported by their mean and standard deviation (std) values across the entire 5.5~h of the independently recorded test set, described in Section \ref{database}.

\begin{table}[t]
	\small
	\renewcommand{\arraystretch}{1.3}
	\caption{Performance without Echo Change.}
	\label{table:aecChNoEchoChange}
	\centering
	\begin{tabular}{| *{7}{c|} }
		\cline{2-7}
		\multicolumn{1}{c|}{}
		& \multicolumn{2}{c|}{UNet}
		& \multicolumn{2}{c|}{Zhang}
		& \multicolumn{2}{c|}{Carbajal}
		\\
		\cline{2-7}
		\multicolumn{1}{c|}{}
		& \multicolumn{1}{c|}{mean}
		& \multicolumn{1}{c|}{std}
		& \multicolumn{1}{c|}{mean}
		& \multicolumn{1}{c|}{std}
		& \multicolumn{1}{c|}{mean}
		& \multicolumn{1}{c|}{std}
		\\
		\hline
		PESQ
		&  \textbf{3.61} &  \textbf{0.24}  &  2.51  &  0.41  &   2.47  &  0.55  \\
		\hline
		SDR
		&  \textbf{7.1} &  \textbf{0.8}  &  4.3  &  1.4  &   4.1  &  1.6  \\
		\hline
		ERLE
		&  \textbf{40.1} &  \textbf{2.1}  &  35.7  &  3.3  &   21.5  &  3.6  \\
		\hline
		SAR
		&  \textbf{8.8} &  \textbf{0.8}  &  4.8  &  1.1  &   4.5  &  1.1  \\
		\hline
	\end{tabular}
\end{table}

\begin{table}[t]
	\small
	\renewcommand{\arraystretch}{1.3}
	\caption{Performance with Echo Change.}
	\label{table:aecChEchoChange}
	\centering
	\begin{tabular}{| *{7}{c|} }
		\cline{2-7}
		\multicolumn{1}{c|}{}
		& \multicolumn{2}{c|}{UNet}
		& \multicolumn{2}{c|}{Zhang}
		& \multicolumn{2}{c|}{Carbajal}
		\\
		\cline{2-7}
		\multicolumn{1}{c|}{}
		& \multicolumn{1}{c|}{mean}
		& \multicolumn{1}{c|}{std}
		& \multicolumn{1}{c|}{mean}
		& \multicolumn{1}{c|}{std}
		& \multicolumn{1}{c|}{mean}
		& \multicolumn{1}{c|}{std}
		\\
		\hline
		PESQ
		&  \textbf{3.3} &  \textbf{0.25}  &  2.35  &  0.45  &  2.05  &  0.7  \\
		\hline
		SDR
		&  \textbf{7} &  \textbf{0.8}  &  2.71  &  1.9  &   2.8  &  1.65  \\
		\hline
		ERLE
		&  \textbf{38.5} &  \textbf{2.45}  &  28.3  &  3.9  &   18  &  4  \\
		\hline
		SAR
		&  \textbf{8.8} &  \textbf{0.95}  &  4.3  &  1.35  &   4.4  &  1.3  \\
		\hline
	\end{tabular}
\end{table}

Results without change in echo path are given in Table~\ref{table:aecChNoEchoChange} and with change in echo path are given in Table~\ref{table:aecChEchoChange}. Our method outperforms competition in all the measures, while also attaining the lowest std. Also, our method is least impeded by the changes in echo path, while the models in \cite{zhang2018deep} and \cite{carbajal2018multiple} both deteriorate in this scenario. Thus, the proposed system provides leading generalization ability to unseen real environments, devices, and speakers and leading robustness to extremely low levels of SER between -10 and -20 dB.

\begin{table}[t]
	\small
	\renewcommand{\arraystretch}{1.3}
	\caption{Performance Before Linear AEC Convergence.}
	\label{table:convergence}
	\centering
	\begin{tabular}{| *{7}{c|} }
		\cline{2-7}
		\multicolumn{1}{c|}{}
		& \multicolumn{2}{c|}{UNet}
		& \multicolumn{2}{c|}{Zhang}
		& \multicolumn{2}{c|}{Carbajal}
		\\
		\cline{2-7}
		\multicolumn{1}{c|}{}
		& \multicolumn{1}{c|}{mean}
		& \multicolumn{1}{c|}{std}
		& \multicolumn{1}{c|}{mean}
		& \multicolumn{1}{c|}{std}
		& \multicolumn{1}{c|}{mean}
		& \multicolumn{1}{c|}{std}
		\\
		\hline
		PESQ
		&  \textbf{2.88} &  \textbf{0.5}  &  2.02  &  0.8  &   1.91  &  0.95  \\
		\hline
		SDR
		&  \textbf{4.9} &  \textbf{1.4}  &  2.6  &  2.1  &   1.1  &  1.7  \\
		\hline
		ERLE
		&  \textbf{31.8} &  \textbf{2.9}  &  23.3  &  4.1  &   15.2  &  4.9  \\
		\hline
		SAR
		&  \textbf{8.5} &  \textbf{1}  &  3.7  &  1.45  &   3.7  &  2.7  \\
		\hline
	\end{tabular}
\end{table}

\begin{table}[t]
	\small
	\renewcommand{\arraystretch}{1.3}
	\caption{Performance for Different Values of $\alpha$.}
	\label{table:alpha}
	\centering
	\begin{tabular}{| *{7}{c|} }
		\cline{2-7}
		\multicolumn{1}{c|}{}
		& \multicolumn{2}{c|}{$\alpha=0$}
		& \multicolumn{2}{c|}{$\alpha=0.5$}
		& \multicolumn{2}{c|}{$\alpha=1$}
		\\
		\cline{2-7}
		\multicolumn{1}{c|}{}
		& \multicolumn{1}{c|}{mean}
		& \multicolumn{1}{c|}{std}
		& \multicolumn{1}{c|}{mean}
		& \multicolumn{1}{c|}{std}
		& \multicolumn{1}{c|}{mean}
		& \multicolumn{1}{c|}{std}
		\\
		\hline
		PESQ
		&  3.61 &  0.24  &  3.54  &  0.29  &   3.45  &  0.35  \\
		\hline
		SDR
		&  7.1 &  0.8  &  6.9  &  0.95  &   6.8  &  1.1  \\
		\hline
		ERLE
		&  40.1 &  2.1  &  41.9  &  2.2  &   43.5  &  2.2  \\
		\hline
		SAR
		&  8.8 &  0.8  &  8.4  &  0.8  &   8.2  &  0.9  \\
		\hline
	\end{tabular}
\end{table}

In the following, we investigate the performance before the linear AEC converges and during re-convergence, in case of changed echo paths. As shown in Table \ref{table:convergence}, performance is collectively impeded when the linear AEC has not converged. However, our method still shows leading performance that points out the high sensitivity of competing methods to converged echo approximation, while the UNet models the residual echo even from degraded measurements.

Next, we demonstrate the effect of $\alpha$ on the tradeoff between RES and desired-signal distortion levels. Again, only unchanged echo paths are considered. Results are presented in Table \ref{table:alpha}. It can be observed that increasing $\alpha$ leads to enhanced RES but at the expense of desired-signal distortion, as suggested by the ERLE and SAR measures, respectively. However, the PESQ, SDR, and SAR measures indicate that for the given $\alpha$ values, the UNet does not severely degrade the quality of the desired signal.

\section{Conclusion}\label{conclusion}
We have introduced an RES method based on a UNet neural network that receives the outputs of a linear AEC in the STFT domain. By using depth-wise separable convolution in the UNet layers, our system consists of 136 thousand parameters that require 1.6 Gflops and 10 MB of memory, which renders it adequate for on-device applications. This system satisfies hands-free communication timing constraints on a standard CPU. In addition, we integrate into the system a tunable tradeoff between echo suppression and signal distortion using a built-in design parameter. Experiments were conducted using 150~h of synthetic and real recordings from the AEC challenge and 11~h of real independent recordings. Results show state-of-the-art performance in real-life conditions in terms of echo suppression and desired-signal distortion compared to competing methods, high generalization to various setups, and robustness to extremely low levels of SERs.

\bibliographystyle{ieeetr}
\bibliography{refs}

\end{document}